# Scientometric Analysis of the German IR Community within TREC & CLEF

*Andreas Konstantin Kruff[1], Philipp Schaer[1]*

1 TH Köln, Germany
{Andreas.Kruff, Philipp.Schaer}@th-koeln.de

**Abstract**

Within this study, the influence of the German Information Retrieval community on the retrieval campaigns Text Retrieval Conference (TREC) and Conference and Labs of the Evaluation Forum (CLEF) between 2000 and 2022 was analyzed based on metadata provided by OpenAlex and further metadata extracted with the GROBID framework from the publication's full texts. The analysis was conducted at the institutional and researcher levels. It was found that the German IR community, both on the author and institution level, mainly contributed to CLEF. Furthermore, it was shown that productivity follows the assumptions made by Lotka's Law.

## 1    Introduction

The scientific field of information retrieval holds international relevance while exhibiting distinctive regional characteristics. This is particularly evident in various evaluation campaigns that have emerged alongside information retrieval conferences, each regionally themed. Notable examples include TREC, which predominantly focuses on English texts, and CLEF, which emphasizes a multilingual approach to prominent European languages. The NII Testbeds and Community for Information Access Research (NTCIR)[1], which specializes in East Asian languages and English, and the Forum for Information Retrieval

---

[1] https://research.nii.ac.jp/ntcir/index-en.html, [last accessed: 02.02.2025]

Evaluation (FIRE)[2] concentrates on Indian languages. Beyond these linguistic distinctions, the area encompasses a variety of domain-specific tasks, as evidenced by evaluation campaigns and the involvement of fields such as question answering, natural language processing, and classification tasks. In this article, the focus will primarily be on examining the German actors in this environment. A similar study has already been done, trying to identify the major contributing institutions and authors from the German Community within the research field of Information Retrieval (Schaer et al., 2023). For that purpose, contributions of German institutes and their corresponding authors to six major IR-related and peer-reviewed conferences, namely CHIIR, CIKM, CLEF, ECIR, SIGIR, and WWW, were further investigated within the period between 2020 and 2023. However, the scope of this work was limited to these six major conferences. The two major evaluation campaigns, namely TREC and the non-peer-reviewed version of CLEF within the CEUR Workshop Proceedings, were not part of the initial analysis. However, it was mentioned that it might lead to the exclusion of "potentially interesting publications and, consequently, maybe the reason for missing research groups." In this work, we will focus the analysis on these two evaluation campaigns specifically and will span the period from 2000 to 2022. The following analysis aims to help identify the key players in terms of German institutions and researchers in the field of Information Retrieval, with a particular focus on the European-North American evaluation campaigns.

## 2    Related Work

While the landscape of analysis in this field, specifically with a focus on the German IR community, is sparse, some works have conducted scientometric analysis in this area.

Baumgartner (2020) analyzed the International Symposium on Information Science (ISI) at the article, author, institutional and reference levels to uncover collaboration patterns, impact, and influence within the field and assess its contribution to the global academic discourse from 1990 to 2004.

Within a similar time frame from 1987 to 1997, a co-word analysis was conducted in the field of IR to uncover established research themes and the rapid change regarding new research themes. However, the work did not focus on the German IR community (Ding, 2001).

---

[2] https://fire.irsi.org.in/fire/2025/home,[last accessed: 02.02.2025]

The same applies to the bibliometric studies done in the context of the two evaluation campaigns in question, which comprised single Tracks or Labs like in Tsikrika et al. (2011) and Thornley et al. (2011) or on a broader level like in Tsikrika et al. (2013) and Larsen et al. (2019). Apart from CLEF and TREC, the impact of IR conference publications like CHIIR was recently studied (Gaede et al., 2024).

Furthermore, the DIRECT Framework was developed, which was capable of identifying and analyzing scientific contributions in the CLEF campaign (Angelini et al., 2014). All those works analyzed the conferences in question, but none of them had a specific focus on the German community in detail.

The most recent work scientometrically analyzing the German IR community was done by Schaer et al. (2023), which takes up methods from the previously mentioned papers and analyzes the most active institutions and researchers within the six major IR-related and peer-reviewed conferences between 2020 and 2023.

In this work, we focus on two key aspects: first, the analysis of the (German) IR community, and second, the techniques and tools employed for article parsing to create a comprehensive framework for our research discussion.

For the parsing, the GeneRation Of BIbliographic Data (GROBID) was used to further enrich the metadata provided by *OpenAlex*. GROBID is a machine-learning library that can convert content from PDF files into XML/TEI files that are easier for machines to process. The implementation is widely used by well-known platforms such as ResearchGate, Semantic Scholar, HAL, and Internet Archive Scholar. Beneficial for this work are the capabilities of parsing names, affiliations and address blocks (GROBID, 2008—2024). According to Tkaczyk et al. (2018), GROBID was demonstrated to be the best-performing out-of-the-box tool for open-source bibliographic reference and citation parsing at that point in time.

## 3  Dataset & Enrichment

The dataset was built upon all publications published since CLEF split off from TREC in 2000 and spans until 2022. However, domain-specific offshots like *TRECVID* were not considered, just like additional special book publications from CLEF. The URL links for CEUR and LNCS publications were scraped from the CLEF initiative website. In contrast, the document links for TREC were scraped from DBLP due to the more uniformly structuring of DBLP compared to the TREC initiative website. Some documents were missing from TREC and were manually added to the dataset. The underlying dataset and the code are made accessible on GitHub[3].

### 3.1 Metadata Enrichment with OpenAlex

In order to analyze the German contributions within the evaluation campaigns, the corresponding metadata needed to be gathered. The reference coverage of OpenAlex competes with that of Web of Science and Scopus, offering their service under a CC0 license, which allows for findings to be reproduced without license barriers (Culbert, 2024). According to Table 1, OpenAlex currently offers the second biggest corpus regarding the number of works compared to services like Dimensions, Web of Science, Scopus, and Google Scholar. While Google Scholar is the only database providing more works, it does not offer a freely available API. Therefore, OpenAlex was chosen as a suitable data resource for the upcoming analysis.

*Table 1: Coverage comparison of scholarly data sources (Source in footnotes [4])*

| Name | # of works (in million) | # OA works (in million) | # Citations (in billion) | Data Openness |
|---|---|---|---|---|
| **OpenAlex** | 243 | 48 | 1.9 | Fully open, CC0 license |
| **Scopus** | 87 | 20.5 | 1.8 | Closed |
| **Web of Science (core)** | 87 | 12 | 1.8 | Closed |
| **Dimensions** | 135 | 29 | 1.7 | Partly open, personal use |
| **Google Scholar** | 389 (estimated) | n.a. | n.a. | Closed |
| **Crossref** | 145 | 20 | 1.45 | Fully open, CC0 license |

---

[3] https://github.com/irgroup/ISI2025-German-IR-Community
[4] https://openalex.org/about#comparison, [last accessed: 10.09.2024]

## 3.2 Metadata Enrichment with GROBID

OpenAlex offers metadata for the geographical location of the institutes and authors that participated in the corresponding work, allowing for further investigation of the German community within this study. However, Table 2 indicates that there is a high discrepancy regarding the coverage of the metadata informations between the three proceedings. While for LNCS, almost every document is provided with this information, for TREC, 35 % are missing, and for CEUR, fewer than half of the documents are provided with it. In order to be able to analyze the German IR community and the top contributing institutions and countries in general, the PDFs were gathered, and the GROBID Framework was applied to overcome those discrepancies. After extracting the metadata information from the PDFs, that could be provided, the coverage regarding the countries and the institutions could be further improved for CEUR and TREC. Table 2 shows that the coverage for the CEUR proceedings could be doubled while the coverage of TREC could be enhanced by ∼15 percentage points. In this case, coverage means that at least one of the two sources provided the metadata. Just the corpus of LNCS could merely improve by 1 %, but it already offered a very high coverage beforehand.

*Table 2: Coverage of metadata information for institutes regarding geographical location (Source: OpenAlex)*

| Proceeding | Coverage before Enrichment in % | Coverage after Enrichment in % |
|---|---|---|
| **CEUR** | 42.23 | 83.00 |
| **TREC** | 65.09 | 80.00 |
| **LNCS** | 95.35 | 96.41 |

Although the coverage could be increased by GROBID, the institution and author names needed to be aligned with the naming conventions of OpenAlex for the following analysis. Due to incomplete extractions by GROBID or lack of uniformity by GROBID and OpenAlex, author and institution names needed to be aggregated. Ambiguities arise for multiple reasons, such as missing diacritical marks, missing middle names, using only the initial of the first name, or other conventions regarding the order of first and last names. The alignment was done manually, but only for the documents with German institutes in-volved since it was sufficient for the analysis. Regarding the institutions, incompletely extracted fragments like the "Animal Sound Archive" were tried to be reassigned to the corresponding institution. Different research teams within one institution and institutions spanning multiple institutional locations with diverse thematic orientations like the Max-Planck-Institute or the Fraunhofer Institute were aggregated to measure the total institutional impact of the institution towards the two evaluation campaigns in question.

# 4   Results

Within this section, we will analyze how much the German IR Community contributed to the two initiatives in question. The analysis will be done on both institutional and individual researcher level. The analysis on the institutional level consists of the analysis of total contributions within the two observed conferences and a descriptive analysis regarding the publication culture, explicitly examining the minimum, maximum, and average number of authors per publication.

The analysis on an individual researcher level also covers the contributions and the interconnectedness of any researcher within the landscape of German institutes. The interconnectedness is expressed and measured based on the implementation of the Betweenness Centrality from the Networkx library[5]. The productivity of the researchers from the German IR community regarding their contribution is further analyzed by applying and comparing it to Lotka's Law.

## 4.1   Analysis on the institutional level

Table 3 gives insight about the input distribution to the total output number from the three proceedings in question. At first, it can be observed that the Top 10 list varies considerably compared to the ranking from Schaer et al. (2023). The only overlap between the two rankings consists of the Webis Group, which is not treated as a single entity in this paper but has been split at an institutional level and is listed in the Top 10 as Bauhaus-Universität Weimar and Universität Leipzig. Furthermore, GESIS also appears in the Top 10 of both rankings. Potential reasons for the discrepancy between the two rankings can be the the-matic orientation of the two conferences towards evaluation campaigns or the differences regarding the observed timeframe.

---

[5] https://networkx.org/, [last accessed: 18.11.2024]

*Table 3: Top 10 most contributing institutes in the German IR Community*

| German Institutes | CEUR | LNCS | TREC | Total |
|---|---|---|---|---|
| **University of Hildesheim** | 31 | 29 | 0 | 60 |
| **University of Hagen** | 23 | 20 | 0 | 43 |
| **DFKI** | 21 | 16 | 2 | 39 |
| **Bauhaus-Universität Weimar** | 11 | 15 | 10 | 36 |
| **Technische Universität Chemnitz** | 22 | 13 | 0 | 35 |
| **RWTH Aachen** | 17 | 17 | 0 | 34 |
| **Universität Leipzig** | 19 | 13 | 1 | 33 |
| **Fraunhofer Institute** | 21 | 4 | 6 | 31 |
| **Humboldt Universität Berlin** | 10 | 11 | 0 | 21 |
| **GESIS** | 9 | 8 | 3 | 20 |

Similar to the ranking of Schaer et al. (2023), the Top 10 shows a strong tendency towards university institutions, except for the DFKI, the Fraunhofer Institute and GESIS. The ranking is clearly characterized by academic non-commercial research institutes and therefore less heterogeneous as the ranking in Schaer et al. (2023).

*Table 4: Number of participating authors from the Top 10 most contributing institutes in the German IR Community*

| German Institutes | Min. # authors | Max. # authors | Avg. # authors |
|---|---|---|---|
| **University of Hildesheim** | 1 | 18 | 4.48 |
| **University of Hagen** | 1 | 7 | 1.88 |
| **DFKI** | 1 | 18 | 4.51 |
| **Bauhaus-Universität Weimar** | 2 | 17 | 7.72 |
| **Technische Universität Chemnitz** | 1 | 11 | 3.17 |
| **RWTH Aachen** | 1 | 17 | 4.59 |
| **Universität Leipzig** | 1 | 17 | 6.73 |
| **Fraunhofer Institute** | 1 | 8 | 2.71 |
| **Humboldt University Berlin** | 1 | 10 | 4.62 |
| **GESIS** | 1 | 16 | 3.95 |

While the University of Hildesheim, the University of Hagen, and the DFKI rank in the top 3 across the total period, a different picture emerges when observing the Top 5 within smaller time frames. Table 6 shows that together with the RWTH Aachen, these three institutes were the most contributing German institutes from 2000-2009. However, from 2010 onwards, the impact of the Bauhaus-Universität Weimar and the Universität Leipzig, which are both part of the Webis Group, kept increasing, making them the two most contributing German institutes in the most recent period. Furthermore, the most recent period is characterized by three institutes, namely the TH Köln, GESIS, and the FHDO, which did not appear in any Top 5 before.

Another observation that can be drawn from the ranking is that all institutions are mainly contributing to the CLEF proceedings and, therefore, show a very homogenous coverage. Only the Bauhaus-Universität Weimar, the Fraunhofer Institute, the University Leipzig, and GESIS contributed works within the TREC community, and only the Bauhaus-Universität Weimar shows a balanced coverage between TREC and CLEF contributions. Together with the Bauhaus-Universität Weimar, the Max Planck Society is the top contributing institute within TREC with ten publications each, followed by the Fraunhofer Institute, the Saarland University, and the B-IT. Except for the Fraunhofer Institute, the coverage between the LNCS and the CEUR proceedings is quite heterogeneous.

The size of the teams involved in the publications of the Top five institutes varied greatly between one and 18, with an average of around four. The Bauhaus-Universität Weimar and Leipzig University stood out with an average of seven to almost eight authors, which could be attributed to their collective identity as the Webis Group, aligning with the findings of Schaer et al. (2023).

The University of Hagen stood out with an average of just under two authors.

## 4.2 Analysis on the Researcher Level

When having a look at the best-connected authors of the German IR community in Table 5, a similar pattern emerged for the institutions concerning participation in TREC challenges. The table shows the Top 10 best connected scientists, that worked for a german institution within the observed time frame. However, the network graph included all the participating authors, that collaborated in a paper, where at least one german institute was involved.

Except for Benno Stein and Matthias Hagen, who both, as part of the Webis Group, contributed nine works within the TREC community only Christoph Friedrich and Martin Potthast contributed a few additional works to TREC. Therefore, eleven of the Top 15 most contributing authors did not contribute within TREC at all.

Here, it could also be observed that most authors contributed almost equally to the two CLEF proceedings, namely LNCS and CEUR. Additionally, it was observed that the total number of output did not automatically contribute to a high betweenness. Hence, Vivien Petras ranked on three with just 13 contributions compared to 43 works of Benno Stein and an almost similar Betweeness Score. Christa Womser-Hacker and Maximilian Eibl contributed more than double the number of publications but ranked sixth and eleven. Therefore, Vivien Petras represented a very important node in the graph by linking multi-ple teams with each other. However, the two most contributing authors, Thomas Mandl and Benno Stein, were also the best-connected authors based on the Betweenness Centrality.

*Table 5: Best-connected authors from German Institutes based on the Betweenness Centrality*

| Authors | Betweenness | Affiliation | TREC | LNCS | CEUR | Total |
| --- | --- | --- | --- | --- | --- | --- |
| **Thomas Mandl** | 0.075 | University of Hildesheim | 0 | 27 | 30 | 57 |
| **Benno Stein** | 0.037 | Bauhaus-Universität Weimar | 9 | 14 | 18 | 41 |
| **Vivien Petras** | 0.036 | Humboldt-Universität Berlin | 0 | 8 | 5 | 13 |
| **Christoph M. Friedrich** | 0.022 | FHDO | 2 | 6 | 12 | 20 |
| **Christa Womser-Hacker** | 0.014 | University of Hildesheim | 0 | 16 | 17 | 33 |
| **Obioma Pelka** | 0.009 | FHDO | 0 | 3 | 6 | 9 |
| **Martin Potthast** | 0.008 | Bauhaus-Universität Weimar | 1 | 14 | 16 | 31 |
| **Bogdan Sacaleanu** | 0.008 | DFKI | 0 | 9 | 9 | 18 |
| **Maximilian Eibl** | 0.007 | TU Chemnitz | 0 | 12 | 21 | 33 |
| **Bayzidul Islam** | 0.007 | TU Darmstadt | 0 | 1 | 2 | 3 |

Figure 1 shows the distribution between the number of authors and the number of publications per author on a log-log scale. The red line describes Lotka's Law. For authors with low publication output, the distribution strictly followed the optimal distribution of Lotka's Law with low variance. It became apparent that there is a small but exceptionally productive core community characterized by the authors in Table 5. With the help of the powerlaw library,[6] the exponent for the given underlying distribution of the German IR Community was estimated as

$$f(n) \propto \frac{1}{n^\alpha} \text{ with } \alpha = 2.63$$

and is, therefore, quite similar to the one assumed by Lotka's Law.

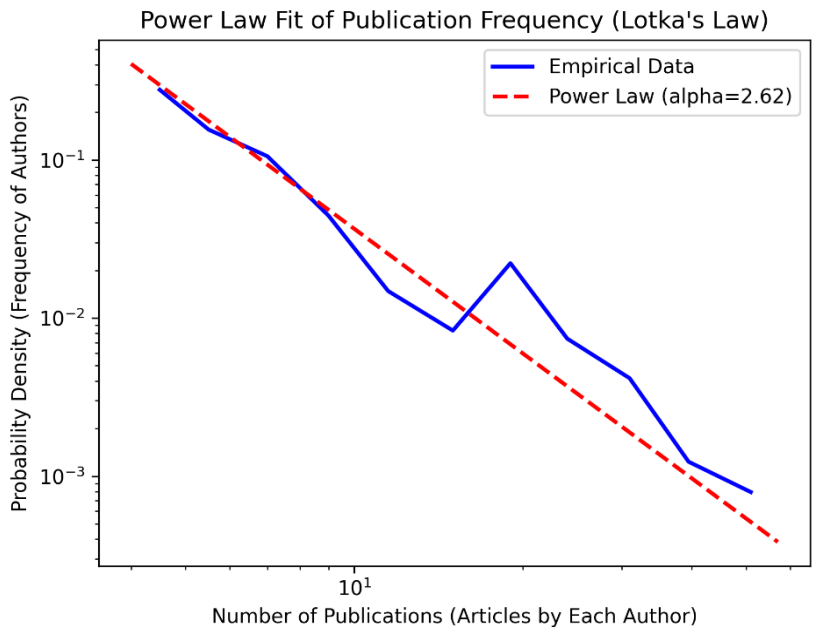

Figure 1: Estimated distribution of the author publication output from the German IR community (created and calculated with the powerlaw library)

## 5 Discussion & Conclusion

### 5.1 Discussion

In this study the contribution of the German IR community in the Evaluation Campaigns TREC and CLEF was analyzed. The in-depth analysis of the Ger-man IR community revealed that with a few exceptions, the top contributing institutions and authors are mainly contributing to the CLEF initiative, and most of them did not contribute to TREC at all. The landscape of the Top 10 contributing institutions of the German IR Community in CLEF and TREC over the last 22 years varies considerably compared to the findings in Schaer et al. (2023). However, some characteristics do align. For example, the Bauhaus-Universität Weimar and the Leipzig University, which are treated separately within this work, show very similar patterns regarding the number of authors compared to their collective identity as the Webis Group in Schaer et al. (2023). While there are a few overlaps regarding the best-connected authors in the German community, the Top 10 vary greatly. This could be related to a shift in time regarding the most relevant authors or, alternatively, different priorities regarding the participation in the evaluation campaigns compared to the proceedings in Schaer et al. (2023).

---

[6] https://pypi.org/project/powerlaw/, [last accessed; 14.11.2024]

At last, it was found that a few authors in the German IR Community are overperforming, when applying Lotka's Law to the authors in the German community. The ratio between the number of authors and number of publications per author aligns very well with Lotka's law, being an indicator of the reliability of the metadata quality of the underlying source. Furthermore, the skewed distribution aligns with the findings of Baumgartner et al. (2020) regarding the distribution of author productivity within the ISI.

## 5.2  Limitations

The biggest limitation of this work is the uncertainty and incompleteness of the bibliometrical data. Not all works could be identified within OpenAlex and some works were falsely aggregated into one entry. This applied especially for works that have been published in both CEUR and LNCS or for consecutive Overview papers from one track.

The poor aggregation and the required data enrichment with GROBID due to the lack of metadata for CEUR, especially and TREC, limited the scope of the community analysis to the German community. The author names needed to be aggregated manually, which would lead to a large magnitude of work doing the analysis on the complete corpus.

Furthermore, the automatic extraction of metadata with the GROBID Framework does not work without errors, which means that potential institutes or names might not always be correctly assigned.

Additionally, in rare cases, OpenAlex's metadata extraction is also prone to errors. In this case, the institutions were wrongly assigned to the given authors, and therefore, some authors appeared in Table 5 and needed to be excluded manually.

### 5.3  Future work

This analysis is only able to capture a selective number of proceedings and, therefore, did not include the analysis of the input regarding contributions from the German IR Community on other IR Evaluation Campaigns like NTCIR or FIRE. This is especially interesting when considering that the analysis has shown how little the German IR community's effect on TREC is.

At the same time, the identified best-connected author, Thomas Mandl, also holds a track chair in FIRE, which would leave the hypothesis open about how well the Asia-Pacific IR communities are connected to the German IR Community.

A more in-depth analysis regarding high betweenness could uncover whether an author's high centrality is attributable to many collaborations with other highly connected researchers or whether the scientist might also have changed teams more frequently.

## Appendix

*Table 6: Top 5 most contributing institutes within the given time periods*

| Time Period | German Institutes | CEUR | LNCS | TREC | Avg. # authors |
|---|---|---|---|---|---|
| 2000-2004 | University of Hildesheim | 6 | 6 | 0 | 3.08 |
| | DFKI | 3 | 2 | 1 | 3 |
| | University of Hagen | 3 | 3 | 0 | 1.33 |
| | RWTH Aachen | 3 | 3 | 0 | 2.33 |
| | Informationszentrum Sozialwissenschaften | 0 | 2 | 0 | 4 |
| 2005-2009 | University of Hildesheim | 19 | 19 | 0 | 4.47 |
| | University of Hagen | 17 | 17 | 0 | 2.03 |
| | RWTH Aachen | 14 | 16 | 0 | 4.87 |
| | DFKI | 11 | 11 | 0 | 4.45 |
| | Technische Universität Chemnitz | 13 | 8 | 0 | 2.33 |
| 2010-2014 | Fraunhofer Society | 8 | 0 | 5 | 3.46 |
| | Technische Universität Chemnitz | 4 | 3 | 0 | 3.43 |
| | Max Planck Society | 4 | 1 | 2 | 4.71 |
| | Bauhaus-Universität Weimar | 1 | 3 | 3 | 5.71 |
| | Humboldt University Berlin | 3 | 3 | 0 | 5.33 |
| 2015-2022 | Bauhaus-Universität Weimar | 10 | 11 | 7 | 8.39 |
| | Universität Leipzig | 16 | 9 | 1 | 7.42 |
| | TH Köln | 6 | 6 | 3 | 3.8 |
| | GESIS | 6 | 4 | 3 | 4.69 |
| | FHDO | 7 | 6 | 0 | 11 |